\documentclass[aps,prl,amsmath,amssymb,twocolumn]{revtex4}
\pdfoutput=1
\usepackage{graphicx}
\usepackage{color}
\newcommand{\beq}{\begin{eqnarray}}
\newcommand{\eeq}{\end{eqnarray}}

\newcommand{\bmp}{\noindent\begin{minipage}{16cm}}
\newcommand{\emp}{\end{minipage}\vskip 7mm} % 7mm untightened

\usepackage{dcolumn}% Align table columns on decimal point
\usepackage{bm}% bold math
\usepackage{bbm}
\usepackage{subfigure}
\usepackage{pxfonts}

\usepackage{epsfig}

\definecolor{rossoCP3}{cmyk}{0,.88,.77,.40}

\def\lsim{\mathrel{\rlap{\lower4pt\hbox{\hskip1pt$\sim$}}
    \raise1pt\hbox{$<$}}}                % less than or approx. symbol
\def\gsim{\mathrel{\rlap{\lower4pt\hbox{\hskip1pt$\sim$}}
    \raise1pt\hbox{$>$}}}                % greater than or approx. symbol

\newcommand{\be}{\begin{eqnarray}}
\newcommand{\ee}{\end{eqnarray}}

\usepackage{fancyhdr}
\newcommand{\drawsquare}[2]{\hbox{%
\rule{#2pt}{#1pt}\hskip-#2pt%  left vertical
\rule{#1pt}{#2pt}\hskip-#1pt%  lower horizontal
\rule[#1pt]{#1pt}{#2pt}}\rule[#1pt]{#2pt}{#2pt}\hskip-#2pt%  upper horizontal
\rule{#2pt}{#1pt}}% right vertical

\newcommand{\Yfund}{\raisebox{-.5pt}{\drawsquare{6.5}{0.4}}}%  fund
\newcommand{\Ysymm}{\raisebox{-.5pt}{\drawsquare{6.5}{0.4}}\hskip-0.4pt%
        \raisebox{-.5pt}{\drawsquare{6.5}{0.4}}}%  symmetric second rank
\newcommand{\Ythrees}{\raisebox{-.5pt}{\drawsquare{6.5}{0.4}}\hskip-0.4pt%
          \raisebox{-.5pt}{\drawsquare{6.5}{0.4}}\hskip-0.4pt% 
          \raisebox{-.5pt}{\drawsquare{6.5}{0.4}}}%  symmetric third rank
\newcommand{\Yasymm}{\raisebox{-3.5pt}{\drawsquare{6.5}{0.4}}\hskip-6.9pt%
        \raisebox{3pt}{\drawsquare{6.5}{0.4}}}%  antisymmetric second rank
\newcommand{\Ythreea}{\raisebox{-3.5pt}{\drawsquare{6.5}{0.4}}\hskip-6.9pt%
        \raisebox{3pt}{\drawsquare{6.5}{0.4}}\hskip-6.9pt
        \raisebox{9.5pt}{\drawsquare{6.5}{0.4}}}

\newcommand{\Yadjoint}{\raisebox{-3.5pt}{\drawsquare{6.5}{0.4}}\hskip-6.9pt%
        \raisebox{3pt}{\drawsquare{6.5}{0.4}}\hskip-0.4pt
        \raisebox{3pt}{\drawsquare{6.5}{0.4}}}%  SU(3) adjoint
\begin{document}
%%%%%%%%%%%%%%%%%%%%%%%%%%%%%%%%%%%%%%%%%%%%%%%%%%%%%%%%%%%%%%%%%%%%%%%%%%%%
%\AddToShipoutPicture*{\BackgroundPic}
\title{\Large  \color{rossoCP3}   Magnetic S-parameter }
\author{Francesco {\sc Sannino}$^{\color{rossoCP3}{\varheartsuit}}$}
\email{sannino@cp3.sdu.dk} 
\affiliation{
$^{\color{rossoCP3}{\varheartsuit}}${ CP}$^{ \bf 3}${-Origins}, 
%IFK \& IMADA, University of Southern Denmark, 
Campusvej 55, DK-5230 Odense M, Denmark.\footnote{CP$^3$- Origins: 2010-25}}
%%%%%%%%%%%%%%%%%%%%%%%%%%%%%%%%%%%%%%%%%%%%%%%%%%%%%%%%%%%%%%%%%%%%%%%%%%%%%%%%%%%%%%%%
%{\flushleft %\begin{figure}
%{\includegraphics[width=3.5cm]{cp3-logo} }}
\begin{abstract}
We propose a direct test of the existence of gauge duals for nonsupersymmetric asymptotically free gauge theories developing an infrared fixed point by computing the S-parameter in the electric and dual  magnetic description. In particular we show that at the lower bound of the conformal window the magnetic S-parameter, i.e. the one determined via the dual magnetic gauge theory, assumes a simple expression in terms of the elementary magnetic degrees of freedom. The results  further support our recent conjecture of the existence of a universal lower bound on the S parameter and indicates that it is an ideal operator for {\it counting} the active physical degrees of freedom within the conformal window. Our results can be directly used to unveil possible four dimensional gauge duals and constitute the first explicit computation of a nonperturbative quantity, in the electric variables, via nonsupersymmetric gauge duality. 
  \end{abstract}

%\end{figure}

\maketitle
 \thispagestyle{fancy}

Quantum chromodynamics (QCD) is the gauge theory describing one of the fundamental forces of Nature, i.e. the one responsible for holding together the quarks inside the proton. {}For over four decades scientists have tried to understand its intricate dynamics using analytical methods as well as first principle computer simulations. Despite the many successes a complete understanding is still missing. The goal of this paper is to shed light on such a complicated dynamics in an innovative way by using a modern version of the Dirac's time-honored idea of electro-magnetic duality to analytically compute nonperturtatbative physically relevant quantities of QCD. Our method is general and can be extended to determine novel nonperturbative quantities for different strongly coupled gauge theories also at nonzero matter density and temperature.

One of the most fascinating possibilities is that generic asymptotically free gauge theories have magnetic duals.
% In fact, in the late nineties, in a series of  ground breaking papers Seiberg  
%\cite{Seiberg:1994pq} provided strong support for the existence of a consistent picture of such a duality within a supersymmetric framework. 
%Using such a duality, Seiberg  has been able to identify the boundary of the conformal window for %supersymmetric QCD as function of the number of flavors and colors.  
Arguably the existence of a possible dual of a generic nonsupersymmetric asymptotically free gauge theory able to reproduce its infrared dynamics must match the 't Hooft anomaly conditions \cite{'tHooft:1980xb}. We have exhibited several solutions of these conditions for QCD and gauge theories with higher dimensional representations respectively in \cite{Sannino:2009qc} and   \cite{Sannino:2009me}. 
%The results were also  in agreement with the all orders beta function  \cite{Ryttov:2007cx}.
% Earlier results can be found in \cite{Terning:1997xy}.
%  are: i) The request that the gauge singlet operators associated to the magnetic baryons should be interpreted as bound states of ordinary baryons \cite{Sannino:2009qc}; ii) The fact that the asymptotically free condition for the dual theory matches the lower bound on the conformal window obtained using the all orders beta function  \cite{Ryttov:2007cx}. These extra constraints help restricting further the number of possible gauge duals without diminishing the exactness of the associate solutions with respect to the 't Hooft anomaly conditions. 

In this work we suggest a direct test of the possible existence of gauge duals using the conformal $S$-parameter \cite{Sannino:2010ca} i.e. the one associated to gauge theories within the conformal window. This parameter is calculable, using the electric theory, near the upper limit of the conformal window \cite{Sannino:2010ca}  since there the electric theory is in a perturbative regime. The results are relevant to shed light on the conformal dynamics and are directly applicable to unparticle extensions of the standard model (SM) \cite{Georgi:2007ek,Sannino:2008nv}. Near the lower boundary of the conformal window we cannot compute $S$ analytically but we expect the magnetic dual to be weakly coupled and hence derive a closed form expression for it via the gauge dual. We will refer to it as the {\it magnetic} $S$ parameter ($S_m$). 

The $S$-parameter \cite{Peskin:1990zt,Peskin:1991sw,Kennedy:1990ib,Altarelli:1990zd}  is  \cite{He:2001tp}:
 \be S&=&-16\pi\frac{\Pi_{3Y}(m_Z^2)-\Pi_{3Y}(0)}{m_Z^2} \,,  
 % T&=&4\pi\frac{\Pi_{11}(0)-\Pi_{33}(0)}{s_W^2c_W^2m_Z^2} \,, 
%  U&=&16\pi\frac{[\Pi_{11}(m_Z^2)-\Pi_{11}(0)]
%              -[\Pi_{33}(m_Z^2)-\Pi_{33}(0)]}{m_Z^2} 
%             \nonumber
\ee
where $\Pi_{3Y}$ is the vacuum polarization of one isospin and one hypercharge current. In the following we use as reference point, instead of the $Z_0$ mass $m_Z$, the external momentum $q^2$.  We couple to the SM a generic gauge theory with sufficient fermionic matter to develop an infrared fixed point (IRFP) with $N_f$ Dirac flavors.   The associated quantum global symmetries are  $SU_L(N_f)\times SU_R(N_f)\times U_V(1)$ if the fermion representation is complex or $SU(2N_f)$ if real or pseudoreal. We weakly gauge $N_D=N_f/2$ doublets. 
To probe the large scale conformal dynamics via $S$, which is UV and IR finite being the difference of the $VV$ and $AA$  two-point functions, we add to the underlying gauge theory a relevant mass operator. This is a standard procedure when trying to investigate the physics of fixed points. We give to the up and down type fermions, with respect to the electroweak interactions an equal mass $m$. The language of the electroweak precision parameters is borrowed to connect more easily to the phenomenological world. 

Having replaced $m_Z^2$ with the momentum $q^2$ the dimensionless $S$-parameter can only be a function of the ratio of $q^2/m^2$. This is so since we assumed the underlying massless gauge theory to be conformal at large distances.  Of course, a dynamical scale is generated when endowing the fermions with masses, however it must be directly proportional to this fermion mass and parametrically smaller. If this were not the case one could never recover the conformal limit when sending the fermion masses to zero. We are henceforth entitled to consider at least two limits with respect to the $q^2/m^2$ ratio  \cite{Sannino:2010ca}: 
%
%
%
%\vskip .2cm
%\noindent
%{\bf Sending $q^2$ to zero keeping fixed the fermion masses:}
%\vskip .2cm
%\noindent
%We assume $M_1=M_2 = m$ and obtain: 
%\begin{eqnarray}
%\lim_{\frac{q^2}{m^2}\rightarrow 0}S =  \frac{\sharp}{6\pi}\left[1 + \frac{1}{10x} + \frac{1}{70 x^2} + {\cal O}(x^{-3})\right] \ , 
%\label{smallq}
%\end{eqnarray}
%with $ x=\frac{m^2}{q^2}$. 
%Note that for $M_1=M_2=m$ the dependence on the hypercharge $Y$ vanishes. We shall be concerned with this limit here but we included the general formulae since they will be useful when considering non-degenerate fermions or when giving different masses only to a certain number of doublets. We observe that, in perturbation theory, when the new gauge interactions are weak, the leading term at zero momentum does not depend on the explicit value of the fermion masses. Turning on a momentum smaller than $m$ we note that the leading dependence is proportional to the inverse of the mass parameter squared. 
%
%
%
%%%
%% \be S&=&-16\pi\frac{\Pi_{3Y}(q^2)-\Pi_{3Y}(0)}{q^2} \,,  
%%%  T&=&4\pi\frac{\Pi_{11}(0)-\Pi_{33}(0)}{s_W^2c_W^2 q^2} \,, 
%%%  U&=&16\pi\frac{[\Pi_{11}(m_Z^2)-\Pi_{11}(0)]
%%%              -[\Pi_{33}(m_Z^2)-\Pi_{33}(0)]}{m_Z^2} 
%%%             \nonumber
%%\ee
%%%
%%where $q^2$ is the square of the external momentum.  $\Pi_{3Y}$ is the vacuum polarization of one isospin and one hypercharge current. Here we use as reference point, instead of the $Z_0$ mass, the external momentum $q^2$. The value assumed by this parameter depends on the specific extension of the SM. 
The one in which the fermion masses go to zero, at finite external momentum and the associated $S$-parameter vanishes and the other one in which the external momentum vanishes first and the  $S$-parameter assumes a nonzero numerical value \cite{Sannino:2010ca}. We have also argued that the latter is the limit which smoothly connects to the $S$-parameter in the chirally broken phase relevant for technicolor. We will therefore concentrate on the limit for $S$ for which  
%\begin{equation}
${q^2}/{m^2} \rightarrow 0$.

%\end{equation}
The {\it electric} $S$-parameter ($S_e$) is defined here as the one computed using the electrical variables. Of course, if the magnetic and the electric theory are gauge duals of each others then $S_m = S_e$. Near the electric (or magnetic) Banks-Zaks \cite{Banks:1981nn}  IRFP this parameter can be computed reliably by means of perturbation theory \cite{Sannino:2010ca} . We found that for an electric $SU(N)$ gauge theory with $N_f$ Dirac fermions transforming according to the representation $r$ of the $SU(N)$ gauge group, and a sufficiently large number of flavors to be near the upper line of the conformal window, the leading terms  in the $q^2/m^2$ expansion and at the leading perturbative order in the gauge coupling constant: 
\begin{eqnarray}
\lim_{\frac{q^2}{m^2}\rightarrow 0}S_e =  \frac{\sharp}{6\pi}\left[1 + \frac{1}{10x} + \frac{1}{70 x^2} + {\cal O}(x^{-3})\right] \ , 
\label{smallq}
\end{eqnarray}
with $ x=\frac{m^2}{q^2}$. 
%The associated quantum global symmetries of the underlying gauge theory are 
% $SU_L(N_f)\times SU_R(N_f)\times U_V(1)$ if the fermion representation is complex or $SU(2N_f)$ if real or pseudoreal.  
 Here  $\sharp = N_D \, d[r] $ counts the number of doublets times the dimension of the representation $d[r]$ under which the fermions transform. {}For example for the fundamental representation $d[F] = N$, for an $SU(N)$ gauge group and $d[S] = N(N+1)/2$ for the two-index symmetric representation of  the gauge group. Note that given that we are in the conformal window the mass to the fermions is given via the standard Higgs mechanism. 

Consider the case of an underlying gauge group $SU(3)$. The
quantum flavor group of the massless theory is 
%\begin{equation}
$SU_L(N_f) \times SU_R(N_f) \times U_V(1)$. 
%\end{equation}
The classical $U_A(1)$ symmetry is destroyed at the quantum
level by the Adler-Bell-Jackiw anomaly. We indicate with
$Q_{\alpha;c}^i$ the two component left spinor where $\alpha=1,2$
is the spin index, $c=1,...,3$ is the color index while
$i=1,...,N_f$ represents the flavor. $\widetilde{Q}^{\alpha ;c}_i$
is the two component conjugated right spinor. We summarize the
transformation properties in the following table.
\begin{table}[h]
\[ \begin{array}{|c| c | c c c|} \hline
{\rm Fields} &  \left[ SU(3) \right] & SU_L(N_f) &SU_R(N_f) & U_V(1) \\ \hline \hline
Q &\Yfund &{\Yfund }&1&~~1  \\
\widetilde{Q} & \overline{\Yfund}&1 &  \overline{\Yfund}& -1   \\
%G_{\mu}&{\rm Adj}   &1&1  &~~0\\
 \hline \end{array} 
\]
\caption{Fermion field content of an SU(3) gauge theory with quantum global symmetry $SU_L(N_f)\times SU_R(N_f) \times U_V(1)$. }
\end{table}
The  global anomalies are associated to the triangle diagrams featuring at the vertices three $SU(N_f)$ generators (either all right or all left), or two 
$SU(N_f)$ generators (all right or all left) and one $U_V(1)$ charge. We indicate these anomalies for short with:
\begin{equation}
SU_{L/R}(N_f)^3 \ ,  \qquad  SU_{L/R}(N_f)^2\,\, U_V(1) \ .
\end{equation}
For a vector like theory there are no further global anomalies. The
cubic anomaly factor, for fermions in fundamental representations,
is $1$ for $Q$ and $-1$ for $\tilde{Q}$ while the quadratic anomaly
factor is $1$ for both leading to
%\begin{equation}
$SU_{L/R}(N_f)^3 \propto \pm 3$, and $SU_{L/R}(N_f)^2 U_V(1)
\propto \pm 3$.
%\end{equation}
 
We have computed the $S$-parameter in the perturbative regime of the conformal window, however we would like now to determine this parameter near the lower bound of the conformal window. Here perturbation theory fails, in the electric variables, and one has to resort to other methods. However, if  a {\it magnetic} gauge dual exists one expects it to be weakly coupled near the critical number of flavors below which one breaks  large distance conformality in the electric variables. 
% This idea is depicted in Fig~\ref{Duality}. 
% \begin{figure}[h!]
%\centerline{\includegraphics[width=8cm]{Duality}}
%\caption{Schematic representation of the phase diagram as function of number of flavors and colors. For a given number of colors by increasing the number flavors within the conformal window we move from the lowest line (violet) to the upper (black) one. The upper black line corresponds to the one where one looses asymptotic freedom in the electric variables and the lower line where chiral symmetry breaks and long distance conformality is lost. In the {\it magnetic} variables the situation is reverted and the perturbative line, i.e. the one where one looses asymptotic freedom in the magnetic variables, correspond to the one where chiral symmetry breaks in the electric ones. }
%\label{Duality}
%\end{figure}
We can then determine $S$ near the lower boundary of the conformal window using perturbation theory in the magnetic variables. 
Determining a possible unique dual theory for QCD is, however, not simple given the few mathematical constraints at our disposal. The saturation of the global anomalies is an important tool but is not able to select out a unique solution. The goal is to find the explicit expression for $S_m$ in terms of  the magnetic variables by means of the most general expectation for the structure of the gauge dual.
 
 As argued in \cite{Terning:1997xy,Sannino:2009qc,Sannino:2009me} a candidate gauge dual theory within the conformal window, saturating the 't Hooft anomaly conditions, would be constituted by an $SU(X)$ gauge group with global symmetry group $SU_L(N_f)\times SU_R(N_f) \times U_V(1)$  featuring 
{\it magnetic} quarks ${q}$ and $\widetilde{q}$ together with $SU(X)$ gauge singlet fermions identifiable as baryons built out of the {\it electric} quarks $Q$. Since mesons do not affect directly global anomaly matching conditions we can add them to the spectrum of the dual theory. In particular they are needed to let the magnetic quarks and the gauge singlet fermions interact with each others. The new mesons will be massless and have no-self potential to respect the conformal invariance of the model at large distances.  We add to the {\it magnetic} quarks gauge singlet Weyl fermions which can be identified with the baryons of QCD but are, in fact, massless. The generic dual spectrum is summarized in table \ref{dualgeneric}.
\begin{table}[h]
\[ \begin{array}{|c| c|c c c|c|} \hline
{\rm Fields} &\left[ SU(X) \right] & SU_L(N_f) &SU_R(N_f) & U_V(1)& \# ~{\rm  of~copies} \\ \hline 
\hline 
 q &\Yfund &{\Yfund }&1&~~y &1 \\
\widetilde{q} & \overline{\Yfund}&1 &  \overline{\Yfund}& -y&1   \\
A &1&\Ythreea &1&~~~3& \ell_A \\
S &1&\Ythrees &1&~~~3& \ell_S \\
C &1&\Yadjoint &1&~~~3& \ell_C \\
B_A &1&\Yasymm &\Yfund &~~~3& \ell_{B_A} \\
B_S &1&\Ysymm &\Yfund &~~~3& \ell_{B_S} \\
{D}_A &1&{\Yfund} &{\Yasymm } &~~~3& \ell_{{D}_A} \\
{D}_S & 1&{\Yfund}  &{\Ysymm} &  ~~~3& \ell_{{D}_S} \\
\widetilde{A} &1&1&\overline{\Ythreea} &-3&\ell_{\widetilde{A}}\\
\widetilde{S} &1&1&\overline{\Ythrees} & -3& \ell_{\widetilde{S}} \\
\widetilde{C} &1&1&\overline{\Yadjoint} &-3& \ell_{\widetilde{C}} \\
M^i_{j} &1&\Yfund &\overline{\Yfund} & 0 &1 \\
 \hline \end{array} 
\]
\caption{Massless spectrum of {\it magnetic} quarks and baryons and their  transformation properties under the global symmetry group. The last column represents the multiplicity of each state and each state is a  Weyl fermion.}
\label{dualgeneric}
\end{table}
The wave functions for the gauge singlet fields $A$, $C$ and $S$ are obtained by projecting the flavor indices of the following operator
\begin{eqnarray}
\epsilon^{c_1 c_2 c_3}Q_{c_1}^{i_1} Q_{c_2}^{i_2} Q_{c_3}^{i_3}\ ,
\end{eqnarray}
over the three irreducible representations of $SU_L(N_f)$ as indicated in the table \ref{dualgeneric}. These states are all singlets under the $SU_R(N_f)$ flavor group. Similarly one can construct the only right-transforming baryons $\widetilde{A}$, $\widetilde{C}$ and $\widetilde{S}$ via $\widetilde{Q}$. The $B$ states are made by two $Q$ fields and one right field $\overline{\widetilde{Q}}$ while the $D$ fields are made by one $Q$ and two $\overline{\widetilde{Q}}$ fermions. $y$ is the, yet to be determined, baryon charge of the {\it magnetic} quarks while the baryon charge of composite states is fixed in units of the QCD quark one. The $\ell$s count the number of times the same baryonic matter representation appears as part of the spectrum of the theory. Invariance under parity and charge conjugation of the underlying theory requires $\ell_{J} = \ell_{\widetilde{J}}$~~ with $J=A,S,...,C$ and $\ell_B = - \ell_D$. 

The simplest mesonic operator is $M_i^{j} $ and transforms simultaneously according to the antifundamental representation of $SU_L(N_f)$ and the fundamental representation of  $SU_R(N_f)$. These states are not constrained by anomaly matching conditions and they mediate the interactions between the magnetic quarks and the gauge singlet fermions via Yukawa-type interactions.  

To probe the chiral properties of the theory requires adding a mass term for the fermions. Near the lower end of the conformal window the dual theory is expected to be weakly coupled yielding the following expression for the magnetic $S$-parameter:
\begin{eqnarray}
S_{m} & = &S_q + S_B +S_{\rm M} \ ,
\end{eqnarray}
with 
\begin{equation}
S_q = \frac{N_D}{6\pi} \, X \ .
\end{equation}
 We will, however, consider here the case in which we gauge, with respect to the electroweak interactions, only the $SU_L (2)\times SU_R (2)$ subgroup where the hypercharge is the diagonal generator of $SU(2)_R$.  In this case only one doublet contributes directly to the $S$ parameter, i.e, we can set $N_D=1$.  This parameter is still sensitive to the whole dynamics. The spectrum of the magnetic quarks, baryons and mesons naturally splits into representations of $SU_L (2)  \times SU_L(N_f - 2) \times   SU_R (2) \times SU_R (N_f - 2) \times U_V(1)$. {}The magnetic quark $q$,  with respect to this group, transforms according to:
\begin{eqnarray}
 q\rightarrow \displaystyle{\left[ (\Yfund,1,1,1)_{y} \oplus (1,\Yfund,1, 1)_{y}\right]} \ .
 \end{eqnarray}
 The baryons have the following decomposition under $SU_L (2)  \times SU_L(N_f - 2) \times   SU_R (2) \times SU_R (N_f - 2) \times U_V(1) $:
\begin{eqnarray}
A & \rightarrow &  {\left[ (1,\Yfund,1,1)_{3} \oplus (\Yfund,\Yasymm,1, 1)_{3}\oplus (1,\Ythreea,1, 1)_{3} \right]}, \nonumber  \\
S & \rightarrow &  \left[ (\Ythrees, 1,1,1)_{3} \oplus (\Ysymm,\Yfund,1, 1)_{3}\oplus (\Yfund,\Ysymm,1, 1)_{3} \oplus \right. \nonumber \\ && , \left.  \oplus (1,\Ythrees,1, 1)_{3}  \right] \nonumber \\
C & \rightarrow &  \left[ (\Yfund, 1,1,1)_{3} \oplus (1,\Yfund,1, 1)_{3}\oplus (\Yfund,\Ysymm,1, 1)_{3} \oplus \right. \nonumber \\ && , \left.  \oplus (\Yfund,\Yasymm,1, 1)_{3}  \oplus (1,\Yadjoint,1, 1)_{3}   \oplus (\Ysymm,\Yfund,1, 1)_{3}  \right] \nonumber \\
B_A & \rightarrow &  \left[ (1, 1,\Yfund,1)_{3} \oplus (1,1,1, \Yfund)_{3}\oplus (\Yfund,\Yfund,\Yfund, 1)_{3} \oplus \right. \nonumber \\ && , \left.  \oplus (\Yfund,\Yfund,1, \Yfund)_{3}  \oplus (1,\Yasymm,\Yfund, 1)_{3}   \oplus (1,\Yasymm,1, \Yfund)_{3}  \right] \nonumber \\
B_S & \rightarrow &  \left[ (\Ysymm, 1,\Yfund,1)_{3} \oplus (\Ysymm,1,1, \Yfund)_{3}\oplus (\Yfund,\Yfund,\Yfund, 1)_{3} \oplus \right. \nonumber \\ && , \left.  \oplus (\Yfund,\Yfund,1, \Yfund)_{3}  \oplus (1,\Ysymm,\Yfund, 1)_{3}   \oplus (1,\Ysymm,1, \Yfund)_{3}  \right]\ . \nonumber \\ &&
\label{split}
\end{eqnarray} 
The decomposition of the charged conjugated baryons is obtained from the one above by exchanging left with right.
% We denote the generic baryon with $\psi^b$ with the label $b=A,S,C, B_A,B_S$. We have demonstrated that $\psi^b$ decomposes according to the spinorial representations $(j^b_1,j^b_2)_{\lambda^b}$ of $SU_L(2)\times SU_R(2)\times U_V(1)$ while the o the left-handed (charge conjugated antiright) baryons $\tilde{\psi}^{\tilde{b}}$ transform according to the representation $(j^b_2,j^b_1)_{-\lambda^b}$.  

Since we are gauging with respect to the electroweak theory the first two flavors we provide a mass term to them as done in \cite{Dugan:1991ck}, i.e. via the introduction of a SM Higgs-type interaction. Since we are operating within the conformal window this is the direct way to provide a mass to the fermions.  By symmetry arguments we can pair only the states which do not transform with respect to $SU_L(N_f - 2) \times SU_R (N_f-2)$ but still transform nontrivially under $SU_L(2) \times SU_R(2)$. These states are $(\Ythrees,1,1,1)_3$ for the baryon $S$; $(\Yfund,1,1,1)_3$ for $C$;  $(1,1,\Yfund,1)_3 $ for $B_A$ and for $B_S$ the state $(\Ysymm,1,\Yfund,1)_3 $. We need to consider the charge conjugated states as well. In terms of the spinorial representations of $SU_L(2)\otimes SU_R(2)$ the states above are ${\ell_S \, (\frac{3}{2},0)_3\oplus \ell_C \, (\frac{1}{2},0)_3\oplus \ell_{B_A} \, (0,\frac{1}{2})_3 \oplus \ell_{B_S} \, (1,\frac{1}{2})_3}$ with the $\ell$ prefactor taking into account the multiplicity of each state. They will pair with their charged conjugated fermion via the mass term operator of the type $\psi H \widetilde{\psi}$ with $H$ the standard model Higgs field which transforms according to the $(\frac{1}{2},\frac{1}{2})$ representation. Note that we can only pair states with $j_2 = j_1 \pm \frac{1}{2}$.
  
Each pair of conjugated fermions transforming according to  $(j_1,j_2)_{\lambda}$ under $SU_L(2)\times SU_R(2)\times U_V(1)$ leads to the following contribution to the $S_m$ parameter  \cite{Dugan:1991ck}: 
\begin{eqnarray}
S_{b} &=& \frac{2\,d_b}{3\pi}\sum_{J J^{\prime}}X_{J,J^{\prime}} \left[ 2f\left(m^2_J,m^2_{J^{\prime}}\right) + g\left(m^2_J,m^2_{J^{\prime}}\right)\right] + \nonumber \\
&+&\left(\frac{\left[ j^- (j^+ + 1)\right]^2}{9\pi}\sum_{J} \frac{2J + 1}{J(J+1)   }\right)  \ , 
\label{sbar}
\end{eqnarray}
with the index $b$ indicating the specific baryon and $d_b$ its degeneracy. We also have $j^- =| j_1 - j_2|$, $j^+=j_1 + j_2$ and $j^- \leq J \leq j^+$ the total spin for each baryon contribution. If more than one spinorial representation belongs to the same baryon $b$ the contributions of all the states must be taken into account. The nonvanishing components of the group theoretical factor $X_{J,J^{\prime}}$ are:  
\begin{eqnarray}
X_{J,J} & = & \left[1- \left(\frac{j^-(j^++1)}{J(J+1)} \right)^2\right] \frac{J(J+1)(2J+1)}{12} \ , \nonumber \\
X_{J, J-1} & = &X_{J-1, J} = \frac{-1}{12} \left((j^+ +1)^2 - J^2 \right)  \left( J^2 - {j^-}^2\right)   \ .
\end{eqnarray}
The functions $f$ and $g$ read \cite{Dugan:1991ck}: 
\begin{eqnarray}
f\left(m^2_J,m^2_{J^{\prime}}\right) &=& -6 \int^1_0 dx\, x(1-x) \log \left(\frac{x m^2_J + (1-x)m^2_{J^{\prime}}}{\mu^2}\right) \ , \nonumber \\
g\left(m^2_J,m^2_{J^{\prime}}\right) &=& 6 \int^1_0 dx \, \frac{x(1-x)m_J m_{J^\prime}}{x m^2_J + (1-x)m^2_{J^{\prime}} } \ .
\end{eqnarray}
The mass of each fermion is directly proportional to the electric fermion mass $m$ and depends on the representation according to the formula 
$m_J = -
%\frac{v}{4\sqrt{2}} 
m \, \frac{J+\frac{1}{2}}{j_1(j_1 +\frac{1}{2})}$. We have chosen as a reference energy scale $\mu =m$. The contribution of  the baryon sector is then:  
\begin{equation}
S_B = \sum_b S_b \ .
\end{equation}

The complex scalar meson $M$ decomposes as:
\begin{equation}
 M \rightarrow \left[(\Yfund,1,\overline{\Yfund},1)\oplus(\Yfund,1,1,\overline{\Yfund})\oplus(1,\Yfund,\overline{\Yfund},1)\oplus(1,\Yfund,1,1,\overline{\Yfund})\right] \ .
 \end{equation} 
 Only the first state,  $(\frac{1}{2},\frac{1}{2})$, contributes to $S_{\rm M}$  and leads to: 
\begin{eqnarray}
S_{\rm M} = \frac{1}{3\pi} \sum_{JJ^{\prime}} f\left(m^2_J,m^2_{J^{\prime}}\right) \ .
\end{eqnarray}
with $J,J^\prime = 1,0$,  $m_J^2 = m^2_0 (1 + J(J+1))$. This is a different mass parameterization than the one given in  \cite{Dugan:1991ck}. We also have $m^2_0 \propto m^2$. All factors of order unity have been set to unity and finally set the scale $\mu = m_0$ in the function $f$ for the scalars.
%The formula in \eqref{sbar} are the ones found in \cite{Dugan:1991ck} generalized here to the magnetic theory. 
The contribution to $S_{\rm M}$ vanishes unless there is  a mass splitting between the different multiplets of the unbroken $SU(2)_V$ symmetry. 

Putting together the various terms we have for the normalized $S_{m}$: 
\begin{eqnarray}
\frac{6\pi}{3}S_{m} = \frac{X}{3} + \frac{\ell_C + \ell_{B_A}}{3} + \frac{25}{729} \, \ell_{B_S} \left( 32 \log2 - 39 \right) - 0.14 \ . \nonumber \\
\label{Sm}
\end{eqnarray} 
The explicit dependence on the quark masses disappear for the $S_m$ parameter in agreement with the expectation from the leading contribution in $q^2/m^2$ to the $S_e$ parameter.  The above is the general expression for $S_m$ near the lower end of the conformal window corresponding to the nonperturbative regime  in the electric variables. {}From this expression is evident that the present definition of the normalized $S$-parameter {\it counts} the relevant degrees of freedom as function of the number of flavors. We estimate $S_m$ using the possible dual provided in \cite{Sannino:2009qc} for which $X = 2 N_f - 15$, $\ell_A = 2$, $\ell_{B_A} = -2 $ (we take $+2$ since we are simply counting the states) with the other $\ell$s vanishing. Asymptotic freedom for the magnetic dual requires at least $N_f=9$ for which $6\pi S_{m}/3 = 1.523$ while if the lower bound of the conformal window occurs for $N_f = 10$ we obtain $6\pi S_{m} /3= 2.19$. Of course, only one of these two values should be considered as the actual value of the normalized magnetic $S$ parameter near the lower end of the electric conformal window. Both values are such that the normalized $S_m$ is always larger than the electrical one near the upper end of the conformal window and are close to the one for two flavors QCD which is around two \cite{Boyle:2009xi}. 
%
%These results support our recent conjecture \cite{Sannino:2010ca}  according to which the normalized $S$ parameter, obtained in the limit when the external momentum vanishes at a nonzero value of the quark mass, is a nondecreasing function of the number of flavors with respect to the underlying electric theory satisfying a universal lower bound corresponding to unity. 
 
The central result \eqref{Sm} rely on the existence of a gauge dual to QCD built extending the famous suggestion of 't Hooft. The form of the dual is general and can be extended to other strongly coupled gauge theories also at nonzero temperature and matter density. Furthermore the existence of a gauge dual can now be finally established by comparing \eqref{Sm} with lattice computations of the same two-point function using the electric variables, i.e. ordinary QCD.

\end{document}